\documentclass{article}
\usepackage{epsfig,amssymb,float,psfig}
\input{epsf}
\newcommand{\be}{\begin{equation}}
\newcommand{\ee}{\end{equation}}
\newcommand{\ba}{\begin{eqnarray}}
\newcommand{\ea}{\end{eqnarray}}
\newcommand{\beq}{\begin{equation}}
\newcommand{\eeq}{\end{equation}}
\newcommand{\beqa}{\begin{eqnarray}}
\newcommand{\eeqa}{\end{eqnarray}}

\newcommand{\ve}{\varepsilon}
\newcommand{\no}{\nonumber}
\newcommand{\vs}{\vspace{-0.275cm}}
\def\barre#1{{\not\mathrel #1}}

\begin{document}


$\,$

\vspace{1cm}

\begin{center}

\bigskip

{{\Large\bf Infrared regularization with spin-3/2 fields
}}

\end{center}

\vspace{.3in}

\begin{center}
{\large
V\'eronique Bernard$^{\dagger,}$\footnote{email: bernard@lpt6.u-strasbg.fr},
Thomas R. Hemmert$^{\ast,}$\footnote{email: themmert@physik.tu-muenchen.de},
Ulf-G. Mei{\ss}ner$^{\ddagger}$\footnote{email: meissner@itkp.uni-bonn.de}
}

\vspace{1cm}

$^\dagger${\it Universit\'e Louis Pasteur, Laboratoire de Physique
               Th\'eorique\\
               F--67084 Strasbourg, France}

\bigskip

$^{\ast}${\it Technische Universit\"at M\"unchen, Physik Department T-39\\ 
              D-85747 Garching, Germany }

\bigskip

$^\ddagger${\it  Universit\"at Bonn, Helmholtz Institut f\"ur Strahlen- 
    und Kernphysik (Theorie)\\ Nu{\ss}allee 14-16, D-53115 Bonn, Germany.}

\bigskip

\end{center}

\vspace{.6in}

\thispagestyle{empty}

\begin{abstract}\noindent
We present a Lorentz--invariant formulation of baryon chiral perturbation theory
including spin-3/2 fields. 
Particular attention is paid to the projection on the spin-3/2 components of the delta fields. 
We also discuss the nucleon mass and the pion-nucleon sigma term.
\end{abstract}

\vfill

\pagebreak

\noindent {\bf 1.} The delta resonance plays a special role in low-energy
nuclear and particle physics, due to its near mass degeneracy with the nucleon
and its strong couplings to pions, nucleons and photons. It was therefore
argued early that spin-3/2 (decuplet) states should be included in baryon
chiral perturbation theory \cite{JMdel}, which is the low-energy effective 
field theory of the Standard Model. The work of \cite{JMdel} and subsequent authors
made use of the heavy baryon approach, which treats the baryons as static
 sources
like in heavy quark effective field theory and allows for a systematic power
counting in the presence of matter fields, as pioneered in \cite{JM} and 
systematically explored in \cite{BKKM}. Also, special care has to be taken about the
decoupling of resonances in the chiral limit \cite{GZ}. This  approach was systematized
by counting the nucleon-delta mass splitting as an additional small parameter in 
Ref.\cite{HHK}, the corresponding power counting was called the  ``small scale
expansion''. The heavy baryon approach has been successfully applied to a
variety of processes, for reviews see \cite{BKMrev,UGM}, and a status report
on chiral effective field theories with deltas is given in \cite{TRH}.
More recently, a Lorentz-invariant formulation of baryon chiral perturbation
theory has become available \cite{BL}, the so--called ``infrared regularization'' (IR).
A similar approach had been presented earlier in \cite{ET}. Such a formulation allows
to include strictures from analyticity and is thus particularly suited for extensions
of chiral perturbation theory based on dispersion relations, see e.g.
\cite{TNT,DGL,GM,OO,OM,AB,bern}. Furthermore, the
use of the Dirac propagator as opposed to the static  fermion propagator in the
heavy baryon scheme allows for a resummation of important recoil effects, see e.g.
the work on electromagnetic form factors \cite{KM}   or the nucleon spin
structure \cite{BHMdhg1,BHMdhg2}. The chiral expansion of the baryon masses
also seems to converge faster in this scheme \cite{ETo} whereas the description
of pion--nucleon scattering is not yet in a satisfactory status \cite{BL2,ETo2}.
The renormalization of relativistic baryon chiral
perturbation theory has been discussed in detail in \cite{FGJS}. In this Letter, we
give a consistent extension of the infrared regularization method in the presence
of  spin-3/2~\footnote{The inclusion of the delta when one separates loop
  integrals into soft and hard parts was already considered in \cite{ETo2}.}.
What is new here is that we explicitly project onto the spin-3/2 parts and
that the method easily allows to include any external source in a chiral
and gauge invariant fashion (in contrast to the recent proposal in \cite{PP}).
As simple applications, we calculate  
the delta loop  contribution to the nucleon self-energy and the 
pion--nucleon sigma term.

\bigskip
\noindent {\bf 2.}
In this section, we briefly review the  formalism necessary to describe spin-3/2 fields.
More details can e.g. be found in \cite{RS,BDM,PT}.
We first write down the propagator for a free spin-3/2 field 
(called the delta propagator from here on) in $d$ space--time dimensions,
\beq\label{fullP}
\mathcal{G}_{\mu\nu}^{\Delta} (p) = -{\barre{p} + m_\Delta \over p^2 - m_\Delta^2}\,
\Biggl\{g_{\mu\nu} - { 1 \over d-1} \, \gamma_\mu \gamma_\nu 
-{ (d-2)\, p_\mu p_\nu \over (d-1)\, m_\Delta^2}
+ { p_\mu \gamma_\nu - p_\nu \gamma_\mu \over (d-1) \, m_\Delta}\Biggr\}~,
\eeq
with $m_\Delta $ the delta mass. This may be rewritten by  
projection onto the spin $3/2$ and $1/2$ components
\beq
\mathcal{G}_{\mu\nu}^{\Delta} (p) 
= -{\barre{p} + m_\Delta \over p^2 - m_\Delta^2}\, P_{\mu\nu}^{3/2}
- {1 \over \sqrt{d-1} m_\Delta} \Biggl( \bigl( P_{12}^{1/2}\bigr)_{\mu\nu} + 
\bigl( P_{21}^{1/2}\bigr)_{\mu\nu} \Biggr) + 
{d-2\over (d-1)\, m_\Delta^2} \, (\barre{p} + m_\Delta )
\, \bigl( P_{22}^{1/2}\bigr)_{\mu\nu}~,
\eeq
with
\beqa\label{proj}
 P_{\mu\nu}^{3/2} &=& g_{\mu\nu} -{1\over d-1}\, \gamma_\mu  \gamma_\nu 
- {1\over (d-1)\,p^2}
\bigl( \barre{p}  \gamma_\mu p_\nu + p_\mu \gamma_\nu \barre{p} \bigr)
 - {d-4 \over d-1} \, {p_\mu p_\nu\over p^2}~,\no\\  
\bigl( P_{12}^{1/2}\bigr)_{\mu\nu} &=& {1\over \sqrt{d-1}p^2} 
\bigl( p_\mu p_\nu - \barre{p}  p_\nu \gamma_\mu)~,\no\\
\bigl( P_{21}^{1/2}\bigr)_{\mu\nu} &=& {1\over \sqrt{d-1}p^2} 
\bigl( \barre{p}  p_\mu \gamma_\nu - p_\mu p_\nu )~,\no\\
\bigl( P_{22}^{1/2}\bigr)_{\mu\nu} &=& { p_\mu \,p_\nu \over p^2}~. 
\eeqa
These spin projection operators fulfill the orthogonality relations
\beq
\left( P^I_{ij} \right)_{\mu\nu} \left( P^J_{kl} \right)^{\nu\rho} =
\delta_{IJ}\,\delta_{jk}\, \left( P^I_{il} \right)^{\rho}_\mu\, , \quad (I,J)
= 1,2 \, , 
\quad (i,j,k,l) = 1,2~.
\eeq
Note the infrared  singular pieces $\sim 1/p^2$ appearing in the spin-projected
parts of the propagator, which will play an important role later on. Also, we remark
that the spin-1/2 pieces do not propagate and thus one should be able to absorb
their contribution in purely polynomial terms (which amounts to a redefinition of
certain low-energy constants in the effective field theory). 
If one considers processes where the center-of-mass energy reaches the delta mass,
$\sqrt{s} = m_\Delta$, the free delta propagator has to be resummed as described
in detail e.g. in \cite{ET,PP}.
Let us now briefly discuss the leading chiral pion-delta-nucleon
Lagrangian. It is given by
\beq\label{Lagr}
{\cal L}_{\pi \Delta N} = c_A \, \bar\Psi_\mu^i \,\,\Theta^\mu_\nu (Z) 
\, w^\nu_i \, \psi
+ {\rm h.c.}~,
\eeq
where $\Psi_\mu$ describes the Rarita-Schwinger (spin-3/2) field, $\psi$ the
nucleon doublet, $w^\mu_i = {\rm Tr}(\tau_i u^\mu)/2$ the axial current  and 
$\Theta_{\mu\nu} = g_{\mu\nu} + (Z-{1/2}) \gamma_\mu \gamma_\nu$. The dependence
on the so-called  off-shell parameter $Z$ is, however, spurious, since it does not appear in
the spin-3/2 contributions and thus physical observables do not depend on it.
The axial-vector coupling constant $c_A$, which is related to the
pion-delta-nucleon coupling
$g_{\pi \Delta N}$, is frequently called $h_A$ in the literature. More precisely,
what appears in Eq.(\ref{Lagr}) is the the  coupling $c_A$ in the chiral limit.
For simplicity, we use in the following the SU(4) coupling constant relation
\beq\label{gpiDN}
c_A = h_A = \frac{3}{2\sqrt{2}}\, g_A~.
\eeq
The resulting value is somewhat larger than  found in typical fits say to
pion-nucleon scattering data or from fitting the $\Delta \to N\pi$ width, 
see e.g.  \cite{FMdel,HHK2}. We remark that like in Eq.(\ref{Lagr}),
we are using standard chiral-invariant couplings of the ${\pi \Delta
N}$ system, because differences to other couplings appearing in the
literature can simply be absorbed in the polynomial contributions to
be discussed in the following.

\bigskip
\noindent {\bf 3.} Next, we make some short remarks about the effective
field theory of massive spin-1/2 fields chirally coupled to Goldstone
bosons and external sources, called baryon chiral perturbation theory. 
It is complicated by the fact that the nucleon mass does not vanish in
the chiral limit and thus introduces a new mass scale apart from the
ones set by the quark masses. Therefore, any power of the quark masses 
can be generated by chiral loops in the nucleon (baryon) case, spoiling the
one--to--one correspondence between the loop expansion and the one
in the small parameter $q$. One method to overcome this is  the heavy mass
expansion (called heavy baryon chiral perturbation theory, for short HBCHPT) 
where the nucleon  mass is
transformed from the propagator into a string of vertices with
increasing powers of $1/m$. Then, a consistent power counting emerges\footnote{The 
extension of this method for including deltas treating the $N\Delta$ mass splitting
as an additional small parameter has been given in \cite{HHK}.}.
However, this method has the disadvantage
that certain types of diagrams are at odds with strictures from analyticity.
The best example is the so--called triangle graph, which enters e.g. the
scalar form factor or the isovector electromagnetic form factors of the
nucleon.  In a fully relativistic treatment,
such constraints from analyticity are automatically fulfilled. It was
argued in~\cite{ET} that relativistic one--loop integrals can be separated
into ``soft'' and ``hard'' parts. While for the former the power counting
as in HBCHPT applies, the contributions from the latter can be absorbed in 
certain LECs. In this way, one can combine the advantages 
of both methods. A more formal and rigorous implementation of such a program 
was given in \cite{BL}. The underlying method  is called 
``infrared regularization''. Any dimensionally regularized
one--loop integral $H$ is split into an infrared singular (called $I$) and 
a regular part (called $R$) by a particular choice of Feynman 
parameterization,
\beq
H = I + R ~.
\eeq
Consider first the regular part. If one  chirally  expands the contributions
to $R$, one generates
polynomials in momenta and quark masses. Consequently, to any order, $R$ can
be absorbed in the  LECs of the effective Lagrangian.  On the other hand, the
infrared (IR) singular part  has the same analytical properties as the full
integral $H$ in the low--energy region and its chiral expansion leads to the
non--trivial momentum and quark--mass dependences of CHPT, like e.g. the
chiral logs or fractional powers of the quark masses.
For a typical one--loop integral (like e.g. the nucleon self--energy $\Sigma$)
this splitting can be achieved in the following way (omitting prefactors)
\beqa
\Sigma = \int  \frac{d^dk}{(2\pi)^d} {1 \over AB} 
&=& \int_0^1 dz \int  \frac{d^dk}{(2\pi)^d} {1 \over [(1-z)A+zB]^2}
\nonumber \\
&=& \biggl\{ \int_0^\infty - \int_1^\infty \biggr\} dz
 \int  \frac{d^dk}{(2\pi)^d} {1 \over [(1-z)A+zB]^2} = I + R~,
\eeqa
with $A=M_\pi^2-k^2-i\epsilon$, $B=m^2 -(p-k)^2 -i\epsilon$,
$\epsilon \to 0^+$,  $M_\pi$ the pion mass and $d$ the number of space--time
dimensions. Any general
one--loop diagram with arbitrary many insertions from external sources
can be brought into this form by combining the propagators to a single
pion and a single nucleon propagator. It was also shown
that this procedure leads to a unique, i.e.
process--independent result, in accordance with the chiral Ward
identities of QCD \cite{BL}.  
Consequently, the transition from any one--loop graph $H$
to its IR singular piece $I$ defines a symmetry--preserving regularization.
For more details, the reader is referred to \cite{BL}.

\bigskip
\noindent {\bf 4.}
Next, we develop a systematic infrared regularization for effective field theories
with spin-3/2 fields. This should be considered the main result of this paper.
Due to the presence of the IR singular terms $\sim 1/p^2$ when one projects
the delta propagator onto its spin-3/2 and spin-1/2 components, cf. Eq.(\ref{proj}),
one has to deal with a type of integrals that do not appear in the pure pion--nucleon
approach. To be specific, consider the integral appearing in the self-energy
(details of the self-energy calculation will be given below)
\beqa\label{J00} 
J_0^0 (p^2) &=& {1\over i} \int \frac{d^4 k}{(2\pi)^4}\, {-1 
\over (k-p)^2(M_\pi^2 -k^2)}\no\\
&=&  \frac{1}{(2\pi )^4}\, \int_0^1 dx \int d^4k \, {1\over [ k^2 + p^2 x
  (x-1) +M_\pi^2 (1-x)]^2}\no\\
&=&  \frac{\pi^d}{(2\pi )^d}\,{\Gamma ( 2 -d/2)\over \Gamma (2)}\, 
\int_0^1 dx  \, {1\over (1-x)^{2-d/2}(-p^2 x + M_\pi^2)^{2-d/2}}~.
\eeqa
In the region of low momenta, where the chiral expansion is expected to converge quickly,
the first factor in the denominator is of ${\cal O}(q^0)$, while the second is of
 ${\cal O}(q^2)$ so that the chiral dimension of the integral should be 
${\cal O}(q^{d-2})$.
We remark that at $x=1$, the integrand develops a pole.
Following \cite{BL}, we perform a change of variables
\beq
x = {M_\pi^2 \over p^2} \, u \equiv \alpha\, u~,
\eeq
so that
\beq
J_0^0 (p^2) = \frac{\pi^{d/2}}{(2\pi )^d}\,{\Gamma ( 2 -d/2)\over \Gamma (2)}\,
{1\over p^2} \,
\int_0^{1/\alpha} du  \, {1\over (1-\alpha u)^{2-d/2}(1-u)^{2-d/2}}\,
(M_\pi^2)^{d/2-1}~.
\eeq
We see that a complication arises due to the infrared singular piece
$\sim 1/(k-p)^2$ in the spin-projected parts of the propagator. Specifically, 
due to the term $\sim\alpha u^2$, the denominator
brings some additional $\alpha$ contribution when performing the integral. 
$J_0^0 (p^2)$ turns out to behave as $M_\pi^{d-4}$ instead of the expected
 $M_\pi^{d-2}$.
If one evaluates this integral straightforwardly, one finds  that it contains a 
logarithmic contribution that indeed has the correct chiral behavior. The disturbing
contribution is due to the factor $(1-\alpha u)$ which stems from the pole at $x=1$
of the infrared singular part of the propagator. 
In fact, such contributions should be contained in the
regular part, since they have nothing to do with the chiral expansion. To formally
achieve this separation, one rewrites the integral as ($d=4-2 \ve$)
\beqa
\int_0^{1/\alpha} du  \, {1\over (1-\alpha u)^{2-d/2}(1-u)^{2-d/2}}
&=& \int_0^{1/\alpha} du  \,(1 -\ve \ln(1-\alpha u) - \ve\ln (1-u)) \no\\
&=& \int_0^{1/\alpha} du  \, ((1 -\ve \ln(1-\alpha u) - 1) +
    \int_0^{1/\alpha} du  \, {1 \over (1-u)^{2-d/2}}~.
\eeqa
From the second  integral one then obtains the irregular part by letting $1/\alpha 
\to \infty$.
Therefore, the separation of the integral ${J_0^0 (p^2)}$ into the irregular and 
the regular terms reads, ${J_0^0 (p^2)}= I_0^0 (p^2) + R_0^0 (p^2)$,
\beqa
I_0^0 (p^2) &=& \frac{\pi^{d/2}}{(2\pi )^d}\,{\Gamma ( 2 -d/2)\over \Gamma (2)}\,
{1\over p^2}\, \int_0^\infty du \, {1\over (1-u)^{2-d/2}}\,(M_\pi^2)^{d/2-1}~, 
\\
R_0^0 (p^2) &=& \frac{\pi^{d/2}}{(2\pi )^d}\,{\Gamma ( 2 -d/2)\over \Gamma (2)}\,
\biggl\{ \int_0^1 dx \,\biggl[ {1\over (1-x))^{2-d/2}} -1 \biggr]-
\int_1^\infty dx \, {1\over (-p^2 x +M_\pi^2)^{2-d/2}}\biggr\}~.
\eeqa
The irregular part can now be worked out using the methodology developed
in \cite{BL}, we find 
\beq\label{IRres}
I_0^0 (p^2) = -{M_\pi^2 \over p^2}\, \Bigl[ 2\bar\lambda +\frac{1}{16\pi^2}
\ln\frac{M_\pi^2}{m^2} \Bigr] = - {\Delta_\pi \over p^2}~.
\eeq
Here, the piece $\sim \bar\lambda$ contains a pole term,
\beq\label{lambdabar}
\bar\lambda = {m^{d-4} \over 16\pi^2} \biggl\{ {1\over d-4} -{1\over 2} \left(
\ln 4\pi + \gamma_E +1 \right) \biggr\}~,
\eeq
with $\gamma_E$ the Euler-Mascheroni constant and we have set the running scale
of dimensional regularization, $\lambda$, to be the nucleon mass \cite{BL}. 
Furthermore, the pion tadpole contribution $\Delta_\pi$ is given by the 
well-known loop integral
\beq
\Delta_\pi = {1\over i} \int {d^d k \over (2\pi)^2}  {1 \over M_\pi^2 - k^2}~.
\eeq
Of course, the
scale that naturally appears in loop integrals with deltas is $m_\Delta$, not $m$,
as used in Eqs.(\ref{IRres},\ref{lambdabar}).
However, the differences due to this choice of scale 
can be absorbed in polynomial (regular) terms, so that
one can work with one scale only in the coupled pion-nucleon-delta system. However, to
assess the theoretical uncertainty at a given order, one can set $\lambda = m_\Delta$
in the delta loop diagrams. Eq.(\ref{IRres}) also shows that the non-propagating
parts of the projected propagator will be absorbed in tadpole contributions,
as it is expected since such contributions can be understood as a very heavy delta
shrunk to a point-like vertex.

\medskip\noindent
Armed with this prescription, we can now set up a consistent expansion in small
external momenta, pion masses and the $N\Delta$ mass splitting $\Delta = m_\Delta - m$, 
in complete analogy to the small scale expansion formulated for heavy spin-3/2
fields \cite{HHK}. We therefore refrain from discussing here in any detail the
corresponding power counting and the construction of the chiral effective Lagrangian
(most of this is available in the literature).
As discussed in detail in \cite{BFHM}, special care has
to be taken to fulfill the decoupling theorem of QCD, which states that leading
chiral singularities can not be modified by resonance fields. From here on,
all small parameters (quark masses, external momenta, nucleon-delta mass
splitting) are collectively denoted as $\epsilon$.

\bigskip
\noindent {\bf 5.}
\begin{figure}[tb]
\centerline{
\epsfxsize=2in
\epsffile{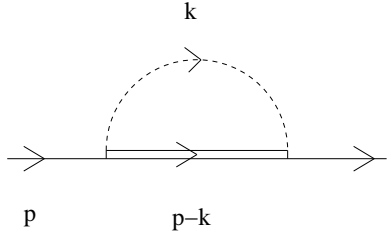}
}
\vspace{0.1cm}
\begin{center}
\caption{Feynman diagram for the nucleon self-energy
with an spin-3/2 intermediate state. Solid, double and dashed
lines denote nucleons, deltas and pions, in order.
\label{fig:self}}
\end{center}
\end{figure}
We first calculate the leading one-loop contribution to the nucleon self-energy 
with an intermediate delta without any reference to a
special regularization and with the full propagator,
Eq.(\ref{fullP}),  see Fig.~\ref{fig:self}. This is done for the following reasons.
First, such a calculation allows one to study different regularization procedures.
Second, one can explicitly study the role of the non-propagating spin-1/2
components. Third, one can also easier compare with existing calculations in the
literature. We have
\beq
\Sigma_\Delta (p^2) = -\frac{9}{4}\frac{g_A^2}{F_\pi^2} \, i\, \int 
\frac{d^4 k }{(2\pi )^4}\,\biggl[ k^\mu - 
\bigl( Z + \frac{1}{2}\bigr)  \barre{k} \gamma^\mu \biggr]\, \mathcal{G}_{\mu\nu}^{\Delta} 
(p-k) \, {1\over M_\pi^2 -k^2} \, \biggl[ k^\nu - \bigl( Z + \frac{1}{2}\bigr) 
\gamma^\nu \barre{k} \biggr]\, , 
\eeq
The general structure of this integral is
\beq\label{selfk}
\Sigma_\Delta (p^2) = {\rm C} \, \frac{1}{i}\, 
\int  \frac{d^4 k }{(2\pi )^4}\, \biggl( A(k) + B(k) \barre{p} + C(k) \barre{k} \biggr)
\eeq
with ${\rm C} = 9g_A^2/(4F_\pi^2)$,
and the functions $A(k)$, $B(k)$ and $C(k)$ can be straightforwardly evaluated.
After integration, the self-energy can be written as
\beq
\Sigma_\Delta (p^2) = \tilde{A} + \barre{p}\, \tilde{B} = 
\tilde{A} +  \tilde{B}\, m + (\barre{p} -m)\,  \tilde{B}~,
\eeq
where the functions $\tilde{A}$ and $\tilde{B}$ follow after integration of
Eq.(\ref{selfk}).
The nucleon mass shift due to the delta intermediate state is 
(setting $Z=-1/2$ for simplicity)
\beqa
\delta m &=& (\tilde{A} +  \tilde{B}\, m)|_{p^2=m^2}\no\\
&=& {(d-2)\,{\rm C}\over (d-1)\,m_\Delta^2} \, 
\biggl( -{1\over 2d}m_\Delta^2 m \Delta_\Delta
- {1\over 2} m_\Delta m^2\Delta_\Delta +{\Delta \over 4} (m_\Delta +m) m
\Delta_\Delta
+ {\Delta \over 2} (m_\Delta +m)^2 m^2 J_1 (m^2) \no \\
&& \qquad\quad +M_\pi^2m^2(m_\Delta+m) J_0(m^2) -{M_\pi^2\over 4} \Delta_\Delta
- m M_\pi^2 J_1(m^2) \biggl[m^2 + {1\over 2}m_\Delta m + {1\over 2}m_\Delta^2\biggr]
\no\\
&& \qquad\quad + {1\over 4} M_\pi^4 m J_1(m^2) + M_\pi^2 \Delta_\pi \biggl[ m_\Delta
+ m + {1\over 2d} m \biggr]\biggr)~,
\eeqa
in terms of the loop functions
\beqa
J_0 (p^2) &=& {1\over i} \int {d^d k \over (2\pi)^2} {1 \over (M_\pi^2 - k^2)
(m_\Delta^2 - (p-k)^2)}~, \no\\
p_\mu \, J_1 (p^2) &=& {1\over i} \int {d^d k \over (2\pi)^2} {k_\mu 
\over (M_\pi^2 - k^2) (m_\Delta^2 - (p-k)^2)}~,\no\\
\Delta_\Delta &=&{1\over i} \int {d^d k \over (2\pi)^2}  {1 \over m_\Delta^2 - k^2}~,
\eeqa
and we have defined 
\beq
\Delta \equiv m_\Delta - m ~.
\eeq
In IR, we have $\Delta_\Delta = 0$, 
and $J_0(m^2)\, (J_1(m^2))$ is
of order ${\cal O}(\epsilon)\, ({\cal O}(\epsilon^2))$, so that to lowest (third) order
\beqa\label{deltamLO}
\delta m &=&  {(d-2)\,{\rm C}\over (d-1)\,m_\Delta^2} \,
\biggl[ \frac{1}{2} \,\Delta \, (m+m_\Delta)^2 \, 
m^2\, J_1(m^2) + M_\pi^2 \, m^2 \, (m+m_\Delta) \, J_0(m^2) \biggr]
\no \\
&=& \!\!\! -{{3g_A^2} \over {16 \pi^2 F_\pi^2}} \biggl[ \Delta \left(\Delta^2-
{3 \over 2}
M_\pi^2\right) \ln{M_\pi^2 \over m^2} + 2 (\Delta^2-M_\pi^2)^{3/2}
\ln{\biggl({{\Delta} \over {M_\pi}}+\sqrt{{{\Delta^2}\over{M_\pi^2}}-1}\biggr)}
 -{\Delta \over 2} \left(\frac{4}{3}\Delta^2-M_\pi^2\right)\biggr]~,\no\\
&&
\eeqa
where the last equation is valid for $\Delta^2 > M_\pi^2$ (as it is the case
in nature). This result agrees with the one in \cite{HHK2}.
Note also that for a more compact notation and easier comparison with the existing
literature, we have retained some higher order terms in the formula expressing the
mass shift in terms of the loop functions $J_0$ and $J_1$.

\medskip\noindent
As noted above, we have not yet separated the spin-3/2 from the spin-1/2 components.
Since the spin-1/2 components do not propagate, their contribution should be entirely
absorbed in polynomial pieces (contact terms). Utilizing the projection operators
given in Eq.(\ref{proj}), the spin-3/2 contribution to the self-energy takes the
form 
\beq
\Sigma_\Delta^{3/2} (p^2) = -\frac{(d-2)\,{\rm C}}{d-1} \, 
\int \frac{d^4 k }{(2\pi )^4}\,
(k^2p^2 - (k\cdot p)^2) { -(\barre{p} + m_\Delta - \barre{k})
 \over (k-p)^2 (M_\pi^2-k^2)
(m_\Delta^2 - (k-p)^2)}~,
\eeq
which leads to the nucleon mass-shift (we give here only the result employing IR)
\beqa\label{delmIR}
\delta m^{3/2}
&=& {(d-2)\,{\rm C}\over (d-1)\,m_\Delta^2} 
\, \biggl[M_\pi^2 m^2 \biggl (J_0(m^2) (m_\Delta+m) -
m J_1 (m^2)\biggr) +  {1\over 2}m^2 \biggl(\Delta (m_\Delta +m) - M_\pi^2\biggr) 
(m_\Delta +m)J_1 (m^2)
 \no \\&& \qquad\quad
+ {1\over 4}\biggl( -\Delta (m_\Delta +m) + M_\pi^2\biggr)^2 m J_1 (m^2) 
\no
- M_\pi^2m^2 \biggl( J_0^0(m^2) (m_\Delta +m) - J_1^0 (m^2) m \biggr)\no\\
&& \qquad\quad + {m^2\over 2} (M_\pi^2 +m^2)(m_\Delta +m) J_1^0 (m^2) -
{m \over 4}  (M_\pi^2 +m^2)^2 J_1^0 (m^2) \biggr]~,
\eeqa
in terms of the new loop functions $J_0^0 (p^2)$ defined in Eq.(\ref{J00}) and
\beq
p_\mu \, J_1^0 (p^2) = {1\over i} \int {d^d k \over (2\pi)^2} {-k_\mu 
\over (p-k)^2 \, (M_\pi^2 - k^2) }~.
\eeq
We also have the loop function relation $J_1^0 (p^2) = (p^2+M_\pi^2)J_0^0 (p^2)
/(2 p^2)  + \Delta_\pi /(2p^2)$ (in dimensional regularization). 
Noting now that the loop functions $J_0^0 (m^2)$
and $J_1^0 (m^2)$ are of order ${\cal O}(\epsilon^2)$ in the chiral expansion, the lowest
(third) order result for the mass shift derived from Eq.(\ref{delmIR}) agrees
with the one given in Eq.(\ref{deltamLO}) but differs of course in the higher
order terms. It is also important to stress that the spin-3/2 contribution to the
mass shift is independent of $Z$, as it should be.

\medskip\noindent
Next, we consider the spin-1/2 contribution to the self-energy,
$\Sigma_\Delta^{1/2} (p^2)$. The corresponding mass shift has the form
(we again give for simplicity the result for $Z=-1/2$)
\beqa
\delta m^{1/2} &=& {3g_A^2\over 2F_\pi^2 m_\Delta^2} {1\over 256} \, M_\pi^4\,
\biggl[ 2\ln {M_\pi^2\over m^2} \biggl( 2(m_\Delta+m) + {2M_\pi^2 \over m^2} 
(2m+m_\Delta) - {M_\pi^4\over m^3} \biggl) \no\\
&&\qquad\qquad\qquad\qquad + {1\over 3}\left( (m+4m_\Delta) + 4{M_\pi^2\over m^2} (2m+m_\Delta)
-2 {M_\pi^4\over m^3}\right)\biggr]~.
\eeqa
We have thus shown explicitly that these contributions can 
be completely absorbed into polynomial
terms appearing in  the chiral expansion of the nucleon mass beyond leading
one--loop order \cite{BM,BL}, 
\beq
m_N = m_0  - 4c_1 M_\pi^2 - {{3g_A^2 M_\pi^3}\over {32\pi F_\pi^2}} +
 k_1 \, M_\pi^4
\ln {M_\pi \over m} + k_2\, M_\pi^4 +
{\cal O}(M_\pi^5)~,
\eeq
where $c_1, k_1$ and $k_2$ are (combinations of) dimension two and four low--energy
constants and $m_0$ is the nucleon mass in the chiral limit. Furthermore, all 
$Z$-dependence is of course included in these polynomial terms, too.

\medskip\noindent
From the mass shift, one can directly derive the  delta loop contribution to the
sigma term using the Feynman-Hellmann theorem,
\beq
\sigma_{\pi N}^\Delta (0) = M_\pi^2 \, {d\delta m \over dM_\pi^2}~.
\eeq
To leading order one has 
(using the full or the spin 3/2 part of the propagator):
\beqa\label{sigmaloop}
\sigma_{\pi N}^\Delta (0) 
&=& \frac{M_\pi^2 (m+m_\Delta)}{m_\Delta^2} \,{\rm C} \, \biggl( m^2 J_0(m^2) + 
\frac{1}{6}\, \frac{1}{4 \pi^2} 
\frac{m^2+2 m_\Delta^2}{m^2}\, \Delta \, (m+m_\Delta)\biggr)\no\\
&=& {{{\rm C} M_\pi^2} \over {4 \pi^2}} \, \biggl[\Delta \, \ln{M_\pi \over m}
 + \sqrt{\Delta^2-M_\pi^2}
\,\ln{\biggl({{\Delta} \over {M_\pi}}+\sqrt{{{\Delta^2}\over{M_\pi^2}}-1}\biggr)
\biggr]}~.
\eeqa
Eq.~(\ref{sigmaloop}) naturally agrees with the result of Ref.~\cite{ETo2} 
but differs by a polynomial contribution in $\Delta$ from the one of 
\cite{ET} due to the use of a $d$--independent propagator in that paper.
Again, for easier comparison we have retained some higher order terms in the
upper expression in Eq.~(\ref{sigmaloop}). 
This concludes the formalism and we now turn to a numerical evaluation of the mass shift
and the sigma term.

\bigskip
\noindent {\bf 6.} We are now in the position to evaluate the delta loop
contribution to the nucleon mass shift and the sigma term. We use $g_A =
1.267$, $M_\pi = 139.57\,$MeV, $F_\pi = 92.4\,$MeV and $\Delta = 271\,$MeV.
From Eq.(\ref{gpiDN}) it follows that $c_A =1.34$.
We remark that the resulting numbers scale
linearly with the axial-vector coupling constant $c_A$.
The delta contribution to the nucleon mass shift and the sigma term are
collected in Table~\ref{tab:res}. We give the lowest (third) order result
which is the same if one uses the full or the projected spin-3/2 propagator,
as pointed out earlier. However, in IR the integrals also contain higher
order pieces which can be retained (see e.g. the detailed discussion in \cite{KM}). 
In that case, one has to project onto
the spin-3/2 pieces to get rid of the unphysical contribution from the
spin-1/2 components. However, as shown in the table, the difference is
numerically irrelevant for the observables considered here. Furthermore,
these higher order corrections amount to a very small correction to the
lowest order result. This, however, is not always the case, see e.g. 
\cite{KM}. We also remark a certain sensitivity to the scale of dimensional
regularization. This dependence, however, will be balanced by counter terms
not considered here.
The results shown in  Table~\ref{tab:res} are not very different
from earlier ones obtained in the heavy baryon approach \cite{J,BKMmass},
which is expected since to this order neither the delta loop contribution to
the nucleon mass shift nor to the sigma term is very  sensitive to recoil corrections 
or similar effects. We stress again that a complete analysis of the nucleon
mass or the sigma term would require a novel determination of certain low-energy
constants to fulfill the requirements of decoupling. This, however,
goes beyond the scope of this Letter.

\renewcommand{\arraystretch}{1.2}
\begin{table}[ht]
\begin{center}
\begin{tabular}{|l|cc|}
\hline
 & $\delta m$ [MeV] & $\sigma_{\pi N}^\Delta (0)$ [MeV] \\
\hline
lowest order                               &   86.4~(108.1)    &  $-$45.6~($-$60.0) \\
unexpanded, with full propagator           &   87.2~(112.7)    &  $-$46.3~($-$63.3) \\
unexpanded, with spin-3/2 propagator       &   88.1~(113.7)    &  $-$44.7~($-$61.4) \\
\hline
\end{tabular}
\caption{\it Delta contribution to the nucleon mass shift and the sigma term.
Lowest order refers to the third order IR result based on 
Eq.(\protect\ref{deltamLO}). Unexpanded means that the higher order terms
from the various loop functions are retained. The propagator used is also
given (in the lowest order case, using the full or the spin-3/2 part of the
propagator leads to the same result, as discussed in the text). The numbers
in the round brackets are obtained with $\lambda = m_\Delta$.
  \label{tab:res}}
\end{center}
\end{table}

\bigskip
\noindent {\bf 7.} In this Letter, we constructed a systematic infrared
regularization for chiral effective field theories including spin-3/2
fields. To arrive at a consistent formulation, one has to project onto
the spin-3/2 components of these fields.
This projection leads to a new type of integrals as compared to the
pure pion-nucleon theory. We have shown that the IR scheme of \cite{BL} can
be extended systematically and furthermore, this procedure allows for a chiral 
invariant inclusion of external sources, like e.g. photons or weak 
currents. The contributions from the spin-1/2 components get
completely absorbed into the polynomial pieces of the effective
Lagrangian. Furthermore, the frequently used off--shell parameters
($X,Y,Z$) are also contained in these polynoms and thus do not lead
to observable consequences.
It is now
important to apply this method to observables, where $\Delta\pi$ loops are
expected to play a significant role, like in all types of Compton scattering.
Work along these lines is underway.

\bigskip\noindent
TRH thanks the Universit\'e Louis Pasteur for hospitality, where part of this
work was done.

\bigskip

\end{document}